# Leptonic Decays of $B$- and $D$-Mesons [*]

R. Sommer[a]

[a]CERN, Theory Division, CH-1211 Geneva 23, Switzerland

The present status of lattice calculations of $f_D$, $f_B$ and some mass splittings are discussed. When one includes the uncertainties due to discretization errors, the results do not yet have a sufficient precision to be relevant to phenomenological applications. There are, however, good prospects of cutting down the uncertainties by a factor of 2 or more soon.

hep-lat/9411024   15 Nov 1994

## 1. INTRODUCTION

$B$-physics plays an important role in the experimental determination of the Cabibbo–Kobayashi–Maskawa matrix and the understanding of CP-violation. Until CP-violation can be observed directly in B-meson decays, hadronic matrix elements are needed in combination with the experimental results on $B - \bar{B}$, $K - \bar{K}$ mixing to obtain restrictions[1] on the famous unitarity triangle and thus on the CP-violating phase $\delta$. Only lattice QCD allows us to compute these matrix elements without model assumptions.

In this review, we do not discuss the relevant $B - \bar{B}$ matrix element but concentrate on the leptonic decay constants $f_B$ and $f_D$, as well as beauty spectroscopy. These quantities are easier to compute. It is important to understand them before one performs a full study of $B - \bar{B}$ mixing.

This talk is an update of ref. [2], which we recommend as an introduction for the non-expert. We restrict our attention to the quenched approximation and refer the reader to ref. [2] for the little that can be said about full QCD.

Compared to last year's conference, where the subject was reviewed by C. Bernard [3], there has not been a rapid development. However, at least in the static approximation, it is well understood how to reduce errors to around the 15% level due to the work of the FNAL group [4] and Draper and McNeile [5] as well as the variance reduction to be discussed in section 5.

---

[*]Invited talk given at the International Symposium on Lattice Field Theory, 27.9–1.10.1994, Bielefeld, Germany

## 2. $f_D$

Leptonic decays, as e.g. $D^- \to l_- \bar{\nu}_l$, are given in terms of a simple vacuum to pseudoscalar matrix element

$$\langle 0 | A_0(0) | P \rangle = f_P \sqrt{m_P/2} \,. \qquad (1)$$

of the relevant axial vector current. In principle, the leptonic decay constant $f_P$ can easily be obtained from a Monte Carlo estimate of the correlator $\langle A_0(x) A_0^\dagger(0) \rangle$.

Nevertheless, there are two non-trivial problems. 1) The axial vector current acquires a renormalization in the lattice regularization and 2) for current values of the lattice spacing $a$, the propagation of the charm quark is distorted because its mass is not small enough compared to the inverse lattice spacing. The second problem becomes a true obstacle for $f_B$.

### 2.1. Renormalization

The relation between the bare lattice current (in the Wilson formulation) and the renormalized current $A_\mu^{f,f'}(x)$ is given by

$$A_\mu^{f,f'}(x) = Z_A(g_0^2, K_f, K_{f'}) \, \bar{q}_f(x) \gamma_\mu \gamma_5 q_{f'}(x), \quad (2)$$

with $K_f$ the hopping parameter of quark flavor $f$ and $g_0^2$ the bare coupling. In non-relativistic normalization and with tadpole improvement [7, 8] it is natural to use

$$A_\mu^{f,f'}(x) = \tilde{Z}_A(g_0^2, K_f, K_{f'}) \sqrt{\frac{1}{2K_f} - \frac{3}{8K_c}} \\ \sqrt{\frac{1}{2K_{f'}} - \frac{3}{8K_c}} \; \bar{q}_f(x) \gamma_\mu \gamma_5 q_{f'}(x) \,. \quad (3)$$

$K_c$ denotes the critical value of the hopping parameter. These two definitions are equivalent when one imposes a non-perturbative normalization condition for the currents, but they are quite different when $Z_A$ and $\tilde{Z}_A$ are taken from 1-loop perturbation theory *neglecting* terms of order $am_f = 1/2K_f - 1/2K_c$. In the numerical estimates, we will use $Z_A$ and $\tilde{Z}_A$ expanded in terms of $\tilde{g}^2 = 3g_0^2/\langle TrU_\square \rangle$, which should be a well-behaved perturbative series [8].

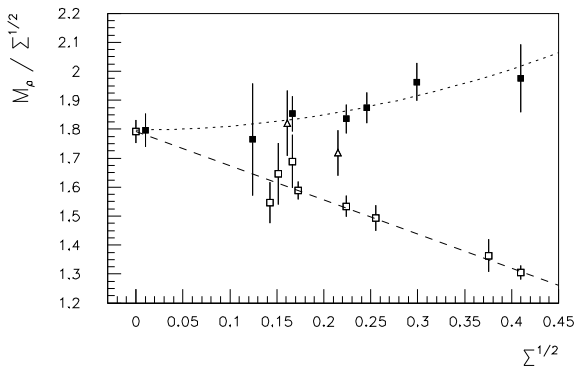

Figure 1. The lattice spacing dependence of $M_\rho/\sqrt{\Sigma}$ for three different fermionic actions: Wilson (squares) SW (triangles) and staggered action (filled squares). $M_\rho$ as calculated by several groups [2] had to be corrected for finite volume effects in some cases. $\Sigma$ is taken from ref. [12] as throughout this review. Note that more recent results [13] at $\Sigma \simeq 0.16$ for the SW-action are lower than the value shown here.

## 2.2. The continuum limit

We use capital letters for quantities in lattice units, e.g. $F_P = af_P$. Since QCD predicts only dimensionless quantities such as mass ratios, we consider the continuum limit of $F_P/F_\pi$. The dominant lattice artefacts should originate from the lowest-dimensional operator which appears in the effective action of lattice QCD but not in the continuum QCD action. For Wilson fermions, this is a dimension-five operator. Hence, we expect the continuum limit to be approached at a rate proportional to $a$ – up to unimportant logarithmic corrections:

$$F_P/F_\pi = f_P/f_\pi + CF_\pi + \dots \quad . \qquad (4)$$

If instead we take the *perturbative* values for $Z_A$ and $\tilde{Z}_A$ and use $M_\rho$ to set the scale, there are additional perturbative corrections

$$F_P/M_\rho = f_P/m_\rho + d\tilde{g}^4 + C'M_\rho + \dots \quad ; \qquad (5)$$

$d$ and higher-order perturbative terms vanish only when the currents are normalized through a non-perturbative normalization condition, e.g. the axial Ward identity [9]. In this case, one still has to remember that $C'$ depends on which normalization condition has been chosen. In fact, for the case of the vector current this $O(a)$ effect is known to be of the order of 15% still at $\beta = 6.4$ [10]. Furthermore, the coefficients $C$ and $C'$ depend significantly on whether one chooses eq. (2) or eq. (3), the difference being of order $\exp(a(m_f + m_{f'})/2)$. We want to point out, however, that in addition to the specific $O(am_f)$ lattice artefacts of heavy mesons, there are sizeable generic order $O(a)$ effects when one uses the standard Wilson action. To show this, we consider the ratio of the $\rho$-mass to the string tension in fig. 1. At $\beta = 5.7$ ($\sqrt{\Sigma} = a\sqrt{\sigma} \simeq 0.4$) there are $\simeq 30\%$ lattice artefacts.

The above general discussion of lattice artefacts serves to underline three points:

- Perturbative corrections are potentially dangerous because they come as powers of $\tilde{g}^2$. Due to their logarithmic variation with the lattice spacing they can hardly be detected numerically. For the case discussed here, they can in principle be avoided by choosing eq. (4).

- There are significant generic lattice artefacts that cannot be eliminated by simply changing the normalization of fields. In order to *reduce* them, one needs to use an improved action and improved currents, the first candidate being the systematic $O(a)$ improvement of Sheikholeslami and Wohlert (SW) [11]. For that action – with tree-level coefficient – there is indirect evidence of improvement compared to the original Wilson action [14]. A direct test of improvement could be obtained through precise computations of $M_\rho/\sqrt{\Sigma}$ in the range $\beta = 5.7 - 6.0$.

- Whichever action and normalization one chooses, the simulation results need to be extrapolated to the continuum using a form like eq. (4). Only in this way can one roughly account for the uncertainty of the continuum results due to $O(a)$ lattice artifacts (that are possibly hidden in the statistical errors).

### 2.3. The present status

An extrapolation of $f_D$ to the continuum limit was reported in ref. [15]. It has been repeated using also the results of other groups [17,19] in ref. [2]. Because of the relatively large statistical errors of $F_\pi$ in these simulations, it is convenient to extrapolate separately $F_\pi/\sqrt{\Sigma}$ and $F_D/\sqrt{\Sigma}$ to the continuum and then take their ratio. In this way, the $O(\tilde{g}^4)$ errors should *roughly* cancel and the results of many groups for $F_\pi$ can be taken to obtain the first ratio. [2]

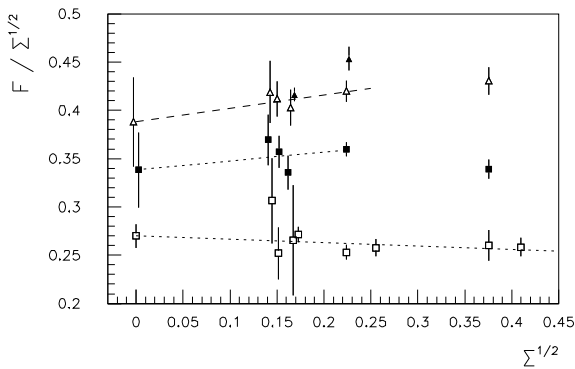

Figure 2. Continuum extrapolation of $F_\pi/\sqrt{\Sigma}$ (squares; data from many groups [2]), $F_D/\sqrt{\Sigma}$ in relativistic normalization (filled squares [15,19,17]) and in non-relativistic normalization (triangles) using $m_2$ as described in the text. For comparison we show also $F_D/\sqrt{\Sigma}$ with the SW-action (filled triangles [16,17]).

As seen in fig. 2, the extrapolation of $F_\pi/\sqrt{\Sigma}$ is rather unproblematic. $F_D/\sqrt{\Sigma}$ does, however,

---

[2] Of course, the results for the decay constants need to be extrapolated in the mass of the light quark and an interpolation in the heavy quark mass needs to be done in order to keep e.g. $M_P/\sqrt{\Sigma} = m_D/\sqrt{\sigma}$ at each value of the lattice spacing.

depend significantly on the normalization of the fields. Compared to the relativistic normalization the data shows much stronger lattice spacing dependence in the non-relativistic normalization, see fig. 4 of ref. [15]. The reinterpretation of the Wilson action as an effective non-relativistic action for large $ma$ suggests [7] to take as the mass of the meson $m_P$ the kinetic mass $m_2$ instead of the pole mass $m_1$ in the non-relativistic expansion of the energy of the meson $E(\vec{p}) = m_1 + \vec{p}^2/(2m_2) + \dots$. Since the simulations mostly do not compute $m_2$, one needs to express $m_2$ in terms of $m_1$ (and possibly the bare quark mass [19]) approximately. This introduces an ambiguity in the procedure, but generally the dependence on the lattice spacing is reduced again. The open triangles in fig. 2 illustrate the weak lattice spacing dependence for the case where one uses the relation between $m_2 a$ and $m_1 a$ that is obtained for a free Wilson quark. We emphasize that these modifications are irrelevant in the continuum limit and one should have results over a large enough range of $a$ such that after the extrapolation the details do not matter any longer.

Inserting the experimental value of $f_\pi$, the above extrapolations yield $f_D = 164(20)$ MeV (relativistic normalization) and $f_D = 188(22)$ MeV (non-relativistic normalization). A safe estimate is therefore $f_D = 176(34)$ MeV

At this conference, two new computations of $f_D$ were reported. We show them in fig. 3 together with previously published values. The preliminary results from the MILC collaboration [20] have very small errors, given the statistical ensemble. If the errors of the final analysis are of this order, one can indeed extrapolate directly $F_D/F_\pi$ and avoid the uncertainties due to the perturbative renormalization. Unfortunately the precision of $F_D/F_\pi$ and the range in lattice spacings where this quantity exists is not sufficient yet to establish its lattice spacing dependence with the SW-action (cf. fig. 3).

It is quite obvious that $f_B$ cannot be computed in this way. A possible approach is given by a non-relativistic treatment of the b-quark [6,7]. In these approaches it is necessary to estimate the systematic errors due to uncertainties in the coefficients in the action and due to the truncation

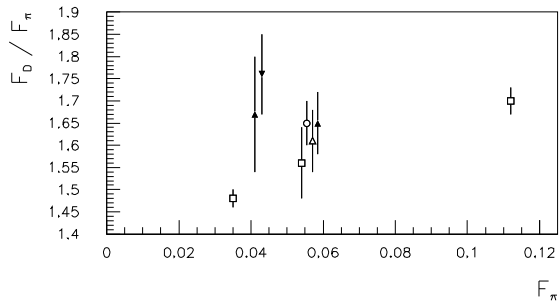

Figure 3. Preliminary results for $F_D/F_\pi$ with the Wilson action (non-relativistic normalization) from ref. [20] (squares), together with the numbers from ref. [21] (diamond) and [17] (triangles). Full symbols are for the SW-action, with the inverted triangle point from ref. [16].

of the action. Applications are presented in [30].

We here discuss the following approach instead. One computes the combination

$$\hat{f} = f_P\sqrt{m_P}[\alpha_s(m_P)/\alpha_s(m_B)]^{2/\beta_0} \quad (6)$$

for a range of masses, say up to $m_p \simeq 3$ GeV. In addition, Eichten's static approximation [22] allows one to determine

$$\hat{f}^{stat} = \lim_{m_P \to \infty} f_P\sqrt{m_P} . \quad (7)$$

In both parts of the calculation, one can take the continuum limit (after renormalizing the current in the static approximation). In the continuum, one then matches the two parts through one or two phenomenologically determined $1/m_p$ correction terms. [3]

## 3. $f_B$ IN THE STATIC APPROXIMATION

A calculation of $\hat{f}^{stat}$ is essential for the success of the above approach. Unfortunately, the static approximation suffers from a bad signal-to-noise ratio of the correlation functions: it is very diffcult to obtain significant correlators beyond a distance of 1/2 fm. In this situation, it is essential to use smearing to suppress the excited states.

---

[3] Such a matching cannot be justified at a finite value of the lattice spacing [19,20]. The reason is that in the regime $am_f \gg 1$ one is essentially in the non-relativistic limit, where one cannot take the continuum limit but must add higher-order operators in order to cancel cutoff effects. Therefore, one may not interchange the interpolation and the continuum limit.

### 3.1. Computing Ground State Properties

Concerning this problem, a true breakthrough has been achieved by the FNAL group [4] and subsequently by Draper and McNeile [5], who showed that the variational approach [23] can be applied successfully to this case. I briefly outline the main principle but *not* the details of the analysis performed by the two groups.

One constructs a matrix of correlators

$$C^{IJ}(x_0) = \sum_{\vec{x}} \langle A_0^{(I)}(x_0, \vec{x})(A_0^{(J)})^\dagger(0)\rangle , \quad (8)$$

$I, J = 0, ..., M$, of axial vector currents with different (Coulomb gauge) wave functions

$$A_0^{(I)}(x_0, \vec{x}) = \sum_{\vec{y}} \bar{q}_{stat}(x_0, \vec{x}+\vec{y})\gamma_0\gamma_5\Phi^{(I)}(\vec{y}) \quad (9)$$

$$q_f(x_0, \vec{x}-\vec{y}) .$$

Its spectral decomposition reads

$$C^{IJ}(x_0) = \sum_{\alpha=0}^{\infty} v_\alpha^I(v_\alpha^J)^* \exp(-E_\alpha x_0) . \quad (10)$$

With $I = 0$ corresponding to the local axial vector current, one has $\hat{F}^{stat} = \sqrt{2}Z^{stat}v_0^0$, where $Z^{stat}$ is the renormalization of the axial current in the static approximation [24]. Diagonalizing the matrix $C^{-1/2}(t_0)C(t)C^{-1/2}(t_0)$ for $t > t_0$, yields estimates $\tilde{v}_\alpha^I$ and $\tilde{E}_\alpha$, which agree with the exact overlaps $v_\alpha^I$ and energies $E_\alpha$ up to corrections of order $O(\exp[-(E_{M+1} - E_\alpha)t])$ [25]. So, one obtains a good estimate of the gap $\Delta E = E_1 - E_0$.

Due to their construction, the trial wave functions $\Phi^{(I)}$ of ref. [4] have a large overlap to the lowest $M$ states. Ref. [5] constructs a *complete* basis of functions $\Phi$. Both groups finally do not take $\hat{F}^{stat}$ from $v_0^0$ directly but use $\tilde{v}_0^I$ to construct smeared–smeared and smeared–local correlation functions to be analyzed in the standard way.

In fig. 4, we show that the results [4,5] are in good agreement. Previous values such as [26] appear rather high (see fig. 20 of ref. [4]). Fig. 5 demonstrates explicitly (using the estimate for the gap $\Delta E$ from ref. [4], and the tables of [26]) that the difference between the results of ref. [5] and [26] is due to an unresolved correction of order $O(\exp(-\Delta E\ t_{min}))$ in the latter reference. (the gap $\Delta E$ not being known, the results were considered as asymptotic within errors at $\exp(-\Delta E\ t_{min}) \simeq 0.3$). Notice in particular that the $O(\exp(-\Delta E\ t_{min}))$ correction becomes larger with decreasing quark mass. Therefore, $f_{B_s}/f_{B_u}$ is strongly affected. In the upper part of the figure, one of the results of [4] receives confirmation from [31]. We conclude that the difference visible in fig. 20 of ref. [4] is quantitatively understood to be due to an excited state contamination.

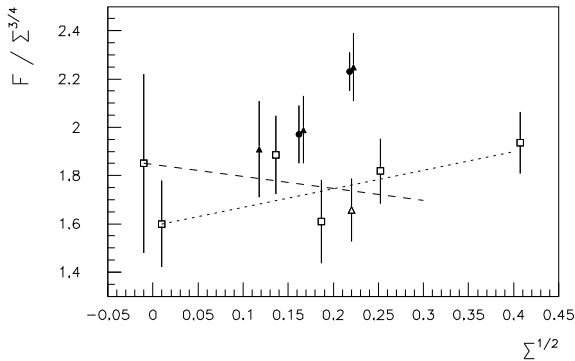

Figure 4. $\hat{F}^{stat}/\Sigma^{3/4}$ with the Wilson action (squares [4], triangle[5]) and the SW action (filled circles [16], filled triangle [18]). Note that for the Wilson action, we use the mean-field improved 1-loop values for $Z^{stat}$ [24]. For the SW action this calculation has not been done yet and the 1-loop results in $\tilde{g}^2$ are used. This is the main reason for the difference in the results.

Ref. [4] investigates carefully also other sources of systematic errors like finite-size effects, and the extrapolation to the physical light quark masses.

### 3.2. The Continuum Limit

Finally, $\hat{F}^{stat}/(\Delta m)^{3/2}$ has been extrapolated [4] to the continuum limit, with $\Delta m$ the 1P–1S charmonium splitting [27]. $\Delta m$ is not known for $\beta = 6.3$. Ref. [4] approximates it by 1-loop evolution from $\beta = 6.1$ to $\beta = 6.3$. This does, however, give an arbitrary value for the $a$-effects of the ratio $\hat{F}^{stat}/(\Delta m)^{3/2}$ (e.g: the same procedure applied also to the numerator simply says that the ratio is independent of the lattice spacing). Fur-thermore, the statistical errors of $\Delta m$ have not been included. We therefore performed the extrapolation of $\hat{F}^{stat}/\Sigma^{3/4}$ using also the data of [5]. The lattice spacing dependence of this ratio is weak. [4] Conservatively, using only the last four points to extrapolate, gives the continuum ratio $\hat{f}^{stat}/\sigma^{3/4} = 1.85(37)$ with an additional (estimated) 7% uncertainty due to the renormalization [24].

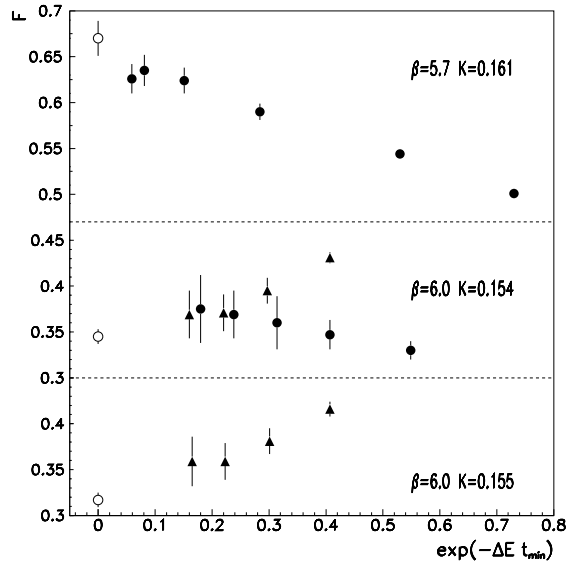

Figure 5. Dependence of the bare $\hat{F}^{stat}$ on the minimum time distance $t_{min}$ in the fit to the local-smeared correlation function. Filled circles correspond to gauge invariant Gaussian smearing functions, while filled triangles are from exponentials [26]. At $\exp(-\Delta E\ t_{min}) = 0$, we show the results of the variational calculations for comparison [4,5]. At $\beta = 5.7$ no fit to the local-smeared correlation function was done, but only one value $t = t_{min}$ was used [31]. The smearing function was not optimized in that case.

In fig. 4, also the available values with the SW action are displayed. Since these calculations do not use the variational method, we suggest that they should be analyzed as shown in fig. 5, to make sure that $O(\exp(-\Delta E\ t_{min}))$ errors are under control. Nevertheless, it is exciting to see that one will soon be able to obtain the continuum limit for this difficult quantity with *two different* actions. The main uncertainty that will remain is the 2-loop uncertainty in the renormalization.

---

[4]Note that figs. 1 and 4 mean that in contrast $\hat{F}^{stat}/M_\rho^{3/2}$ decreases strongly with decreasing $a$, explaining the very high numbers that were obtained originally [28,29].

**4.** $f_B$

One may now combine $\hat{f}^{stat}/\sigma^{3/4}$ with $m_\rho/\sqrt{\sigma}$ or $f_\pi/\sqrt{\sigma}$ to obtain $\hat{f}^{stat}$ in physical units. Together with $\hat{f}(m_P)$ in the range $1.2 \text{ GeV} \le m_P \le 3 \text{ GeV}$ obtained as in sect. 2.2, one finally determines $f_B = 180(48)$ MeV through interpolation. First results using the actions [6,7] are in agreement with this value [30].

A significant improvement of this result should be possible. Reducing the lattice spacing will allow for a more precise extrapolation of $\hat{f}(m_P)$ and a wider range of $m_P$. At the same time, it is necessary to improve the statistical accuracy, especially in the static approximation. In the following section, we demonstrate that a significant factor can be obtained without additional computational effort.

## 5. VARIANCE REDUCTION

Consider for simplicity the correlation function of two pseudoscalar densities. After integration over the Grassmann variables, it can be written

$$O_1(x_0) = \sum_{\vec{x}} \langle Tr S_f(x_0, \vec{x}; 0, \vec{y}) S_{f'}^\dagger(x_0, \vec{x}; 0, \vec{y}) \rangle,$$

where the average $<>$ is over the gauge fields with the appropriate weight including possibly the fermion determinant and $S_f$ is the quark propagator of flavor $f$. The variance of this correlation function could be decreased by averaging over $\vec{y}$. Straightforwardly this is not possible, since it requires quark propagators to be calculated from each point $\vec{y}$. Instead, with just the effort necessary to compute $S$ from one point $\vec{y}$, we can calculate the combination $\bar{S}_f(x_0, \vec{x}) \equiv \sum_{\vec{y}} S_f(x_0, \vec{x}; 0, \vec{y}) \sigma_{\vec{y}}$ where $\sigma_{\vec{y}}$ is a random field of Ising variables. With this building block, a second observable ($N \equiv \sum_{\vec{y}} 1$)

$$O_2(x_0) = \frac{1}{N} \sum_{\vec{x}} \langle Tr \ \bar{S}_f(x_0, \vec{x}) \bar{S}_{f'}^\dagger(x_0, \vec{x}) \rangle$$

can be constructed, which on average is exactly equal to $O_1$. Here it is understood that we average also over the Ising field. The variance of $O_2$, $V(O_2)$, involves in addition to the terms that are present in $V(O_1)$ 4-point functions that are partly summed over their arguments. If these 4-point functions decay fast enough as the arguments are separated, $V(O_2) = O(1/N) \times V(O_1)$. The prefactor may however also be large, such that on a lattice of $(1.5 \text{ fm})^3$ one has $V(O_2) > V(O_1)$ because $N$ is not large enough. This prefactor originates from the sum over short distances in the 4-point functions. It can be decreased by not summing over every point $\vec{y}$, but over "well separated" points only.

The essential idea to reduce the variance of 2-point functions like this was given in ref. [32]. There, and in a recent investigation [33], it was concluded that the method does not improve the variance in the cases of practical interest. However, we do not expect a large prefactor when at least one quark flavor is heavy.

This variance reduction can be applied for any action of the quarks and for any type of smearing.

The idea has been tested both in the static approximation and for a heavy quark around the charm-mass at $\beta = 5.7$ on a $12^3 \times 24$ lattice and with smearing [31]. As suggested by the above argument, one finds a shallow minimum of $V(O_2)$ as a function of the separation between the points $\vec{y}$. The minimum occurs around a separation of 0.3 fm [31]. It is considerably lower than $V(O_1)$: $V(O_1)/V(O_2) = 3.4 - 4.5$ in the relevant time interval in the static approximation and $V(O_1)/V(O_2) = 2.6 - 4.0$ for a charm–light correlator. One clearly expects that these ratios will grow proportionally to the space-like volume, when the separation of points is kept fixed. It was also checked that the gain translates e.g. into a factor $1/2.5$ in the error on $\hat{F}^{stat}$ on the $12^3 \times 24$ lattice.

Such factors should not be missed in future heavy-light and heavy-heavy computations.

## 6. BEAUTY/CHARM SPECTROSCOPY

The spectroscopy of mesons and baryons with b- or c-content, is of twofold interest. On the one hand, there are still channels where lattice gauge theories can make predictions; on the other hand, one can check the importance of systematic errors like the quenched approximation against experimental numbers and one can test the size of $1/m$ corrections to the heavy quark limit in these quantities. Two splittings have been studied systematically.

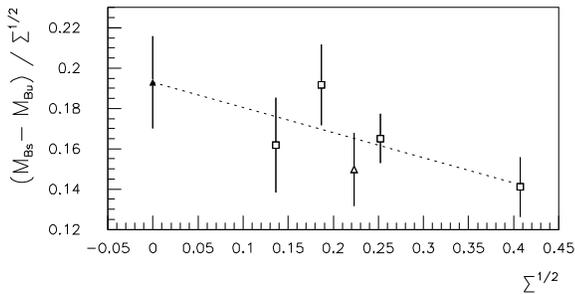

Figure 6. Continuum extrapolation of $(M_{B_s} - M_{B_u})/\sqrt{\Sigma}$. Data are from [4] (squares) and [13] (triangle).

The splitting $M_{B_s} - M_{B_u}$ was computed in the static approximation for a range of lattice spacings by ref. [4]. My remark concerning the continuum limit extrapolation in sect. 3.2 applies here as well. I have done the extrapolation of $(M_{B_s} - M_{B_u})/\sqrt{\Sigma}$, see fig. 6. I have included the result of [13] in this case, since it is well known that the excited state contributions are much less relevant for the extraction of the masses than for the determination of the decay constants. Indeed, a preliminary value [5] obtained with the variational technique is in agreement with ref. [13].

In the continuum limit one obtains $(m_{B_s} - m_{B_u})/\sqrt{\sigma} = 0.19(2) + O(\sqrt{\sigma}/m_B)$. This is to be compared to the "experimental" result $(m_{B_s} - m_{B_u})/\sqrt{\sigma} = 89(4)$ MeV$/(420 - 440)$ MeV $= 0.20(1) - 0.21(1)$, where the real uncertainty in assigning a value to the string tension is hard to quantify. Unless there is an accidental cancellation of the two effects, the influence of dynamical fermions and the $O(1/m_b)$ terms are not very important. This is in agreement with the experimental fact that the splitting $m_{D_s} - m_{D_u}$ is only $\sim 10\%$ higher.

$M_{\Lambda_b} - M_B$ has been computed with the Wilson action for a range of lattice spacings with varying masses of the heavy quark [34]. The lattice spacing effects of $(M_{\Lambda_b} - M_B)/M_\rho$ are small. One can extrapolate to the continuum and to the mass of the b-quark [34]. In the static approximation there are estimates at $\beta = 5.74$ [26] and at $\beta = 6.0$ [35] which, however, are obtained from only moderately long plateaus. Stella has presented results from UKQCD obtained at $\beta = 6.2$ with the SW action [36]. They are in nice agreement with [34], thus indirectly confirming the smallness of lattice spacing effects in $(M_{\Lambda_b} - M_B)/M_\rho$. The consistent picture that emerges for this splitting is summarized in a plot of a poster presented by Borrelli [37] at this conference. The mass dependence of this splitting is again rather weak, suggesting only an $\sim 15\%$ change between the b- and the c-mass.

For the promising investigation of a number of other splittings, I refer the reader to [36].

## 7. CONCLUSIONS

A significant advance has been made in the static approximation by applying the variational technique [4,5] to obtain ground state properties. Not only do refs. [4,5] obtain reliable numbers for the decay constants, but with the help of the gap computed in ref. [4] one can quantitatively estimate the contamination due to the first excited state that is present in other calculations. Soon, the precision of $f^{stat}$ may be limited mainly by the unknown 2-loop effects in the renormalization.

Concerning the computations at finite mass, we point out that it is not sufficient to correct (approximately) for one type of $O(a)$ effect or another. One needs to perform calculations over a range of lattice spacings with *one* action, *one* normalization of the fields, *one* definition of the meson mass and extrapolate to the continuum. The action of ref. [7] and/or the SW action should help in that they may allow for a smoother continuum extrapolation than the standard Wilson action. We can also learn more about $O(am_f)$ effects once the 1-loop calculations of ref. [7] are finished.

Higher-precision calculations are under way. I hope that by the time of the next conference the final errors on $f_D$ and $f_B$ can be cut by about a factor 2 or more, especially if the variance reduction described in section 5 is applied.

In this review, we combined data from different groups to perform continuum extrapolations of certain quantities. As the reference scale we used the string tension because it is known with reasonable precision for the relevant range of $\beta$-values. A related quantity, $r_0$, is known to be much better for this purpose [38]. Once it will have been computed, we will not need to worry about the residual systematic errors in the determination of the string tension.

In order to finally compute the $B - \bar{B}$ mixing amplitude, the $\Delta b = 2$ four-fermi operator needs to be renormalized. It remains a true challenge to perform this renormalization non-perturbatively or "at least" to two loops.

**Acknowledgement** I thank all my colleagues who, through discussions, helped me learn about this subject. I thank S. Güsken for a critical reading of the manuscript.